%% file: mdwf_fk_fpi.tex
\documentclass[aps,prd,twocolumn,tightenlines,preprintnumbers,showpacs,superscriptaddress,notitlepage,nofootinbib,floatfix,longbibliography,10pt]{revtex4-1}

\input{preamble}

\input{def}

\input{affiliations}

\newcount\hour \newcount\hourminute \newcount\minute
\hour=\time \divide \hour by 60
\hourminute=\hour \multiply \hourminute by 60
\minute=\time \advance \minute by -\hourminute

\begin{document}

\title{
$F_K/F_\pi$ as a precision test of a new four flavor Domain Wall Fermion action
}

\author{Renwick J. Hudspith}
\affiliation{\cmu}

\author{Nicolas Garron}
\affiliation{\liverpoolHope}
\affiliation{\liverpool}

\author{Zack Hall}
\affiliation{\lblnsd}

\author{Andrew Hanlon}
\affiliation{\kentState}

\author{Henry Monge-Camacho}
\affiliation{\ornl}

\author{Colin~Morningstar}
\affiliation{\cmu}

\author{Amy Nicholson}
\affiliation{\unc}

\author{Dimitra A. Pefkou}
\affiliation{\ucb}
\affiliation{\lblnsd}

\author{Thomas R.~Richardson}
\affiliation{\ucb}
\affiliation{\lblnsd}

\author{Fernando Romero-L\'{o}pez}
\affiliation{\bern}

\author{Miguel~Salg}
\affiliation{\bern}

\author{Wyatt~A.~Smith}
\affiliation{\wm}
\affiliation{\lblnsd}
\affiliation{\ucb}

\author{Pavlos Vranas}
\affiliation{\llnl}
\affiliation{\lblnsd}

\author{Andr\'{e} Walker-Loud}
\affiliation{\lblnsd}
\affiliation{\ucb}

\author{Bigeng Wang}
\affiliation{\kentucky}
\affiliation{\lblnsd}

\date{\today}

\begin{abstract}
We present a new set of lattice QCD ensembles with four flavors of smeared M\"obius Domain Wall Fermions with good chiral symmetry and small fifth-dimensional extent. A modest amount of computing resources was sufficient to generate 30 publicly-available ensembles spanning five lattice spacings and a broad range of pion masses down to physical. To scrutinize our action we determine $F_{K^{\pm}}/F_{\pi^{\pm}} = 1.1962(34)$, a key quantity for precision CKM unitarity tests, heralding a future of inexpensive high-precision calculations of hadronic observables with chiral fermions.
\end{abstract}

\maketitle


\input{introduction.tex}

\bigskip
\input{ensemble_details.tex}

\bigskip
\input{chiral_fit_form.tex}

\bigskip
\input{results.tex}

\bigskip
\input{conclusions.tex}

\bigskip\noindent\textbf{DATA AVAILABILITY:}
The correlation functions and bootstrap samples of the ground state masses and overlap factors are available at
\url{https://portal.nersc.gov/cfs/m2986/cosmon/mdwf/fk_fpi_2026/}.
The \texttt{mdwf} branch of this git repository~\cite{callat_fkfpi} can be used to reproduce the extrapolation analysis presented here.
The gauge fields used in this work are all available via Globus.
From the Globus ``FILE MANAGER'', search for the collection ``NERSC Gauge Connection'' and then they can be found in the directory \texttt{cosmon/mdwf/nf211\_8stout}.
Contact us if you have any difficulty accessing the data or configurations.

\acknowledgments

We would like to thank Vincenzo Cirigliano for help with the first row CKM unitarity analysis
and Takeshi Kaneko for help understanding the FLAG averaging procedure and Peter Boyle for discussions about using Grid.

Ensemble generation was performed in a modified version of \verb|Grid| \cite{Boyle:2015tjk}. Pure gauge quantities were computed with \verb|GLU| \cite{Hudspith:2014oja}. M\"obius Domain Wall propagator inversions for meson correlation functions were done with \verb|QUDA| \cite{Clark:2009wm,Babich:2011np}, while contractions were performed in \verb|QDP++| \cite{Edwards:2004sx}. Fits to the data were partly performed using the libraries \verb|gvar|~\cite{gvar} and \verb|lsqfit|~\cite{lsqfit}. This research used resources of the National Energy Research Scientific Computing Center (NERSC), a Department of Energy User Facility using NERSC awards NP-ERCAP[0033436,0032885,0036425,0036336],
and the Carnegie Mellon University QCD cluster~\cite{cmu_cluster}.

WAS acknowledges support from startup funds at William \& Mary provided to Andrew Jackura. 
The work of FRL and MS was supported in part by the Swiss National Science Foundation (SNSF) through grant No. 200021-236432.
This work was supported in part by the U.S. National Science Foundation (NSF) under grant OAC-2311430 (RJH, CM, AWL),
through the Mathematical and Physical Sciences Ascending Postdoctoral Research Fellowship under award No. 2402482 (ZH), under NSF Award PHY-2514831 (CM), by the NSF through cooperative agreement 2020275 (TRR), and under the NSF Faculty Early Career Development Program (CAREER) under award PHY-2047185 (AN);
This work was supported in part by the U.S. Department of Energy (DOE), Office of Science, Office of Nuclear Physics under grant contract numbers DE-AC02-05CH11231 (ZH, DAP, TRR, WAS, AWL, BW), 
DE-SC0013065 (BW),
by Lawrence Livermore National Laboratory (LLNL) under Contract DE-AC52-07NA27344 (PV),
through the “Nuclear Theory for New Physics” Topical Collaboration award No. DE-SC0023663 (ZH, CJM, DAP, TRR, AWL),
through Scientific Discovery through Advanced Computing (SciDAC) award ``Computing the Properties of Matter with Leadership Computing Resources'' (AWL).
This research used resources of the Oak Ridge Leadership Computing Facility at the Oak Ridge National Laboratory, which is supported by the Office of Science of the U.S. Department of Energy under Contract No. DE-AC05-00OR22725 (HMC).

This  manuscript  has  been  authored  by  UT-Battelle,  LLC,  under  Contract No.  DE-AC0500OR22725  with  the  U.S.  Department  of  Energy.  The  United States  Government  retains  and  the  publisher,  by  accepting  the  article  for publication, acknowledges that the United States Government retains a non- exclusive, paid-up, irrevocable, world-wide license to publish or reproduce the published  form  of  this  manuscript,  or  allow  others  to  do  so,  for  the  United States Government purposes. The Department of Energy will provide publicaccess to these results of federally sponsored research in accordance with the DOE Public Access Plan.

\bibliography{bib}

\clearpage

\noindent
\input{end_matter}

\end{document}

%% file: preamble.tex
\usepackage{graphicx}  
\usepackage{dcolumn}   
\usepackage{bm}        
\usepackage{amssymb}   
\usepackage{standalone}
\usepackage{enumitem}
\usepackage[pdftex]{color}
\usepackage[utf8]{inputenc}
\usepackage{xcolor}
\usepackage{slashed}
\usepackage{booktabs}
\usepackage{multirow}
\usepackage{amsmath}
\usepackage{bbm}
\usepackage{stackrel}
\usepackage{rotating}
\usepackage{CJKutf8}
\usepackage{pifont} 
\newcommand{\cmark}{\textcolor{green}{\ding{51}}}%
\newcommand{\xmark}{\textcolor{red}{\ding{55}}}%

\usepackage{mathtools}
\usepackage[caption=false]{subfig}
\usepackage{listings}

\usepackage{hyperref}
\hypersetup{
    colorlinks=true,       
    linkcolor=blue,          
    citecolor=blue,        
    filecolor=blue,      
    urlcolor=blue           
}
\usepackage{simplewick}
\usepackage{float} 
\usepackage{tikz} 
\usepackage{xspace}
\usepackage[normalem]{ulem}

\usepackage{comment}

\allowdisplaybreaks

\AtBeginDocument{%
    \newwrite\bibnotes
    \def\bibnotesext{Notes.bib}
    \immediate\openout\bibnotes=\jobname\bibnotesext
    \immediate\write\bibnotes{@CONTROL{REVTEX41Control}}
    \immediate\write\bibnotes{@CONTROL{%
    apsrev41Control,author="08",editor="1",pages="1",title="0",year="1"}}
     \if@filesw
     \immediate\write\@auxout{\string\citation{apsrev41Control}}%
    \fi
}%

%% file: def.tex
\def\eqref#1{{(\ref{#1})}}
\newcommand{\eqnref}[1]{Eq.~\eqref{#1}}

\newcommand{\figref}[1]{Fig.~\ref{#1}}

\newcommand{\tabref}[1]{Table~\ref{#1}}

\definecolor{kngrey}{HTML}{A6AAA9}
\definecolor{knred}{HTML}{EC5D57}
\definecolor{knorange}{HTML}{F39019}
\definecolor{knyellow}{HTML}{F5D328}
\definecolor{kngreen}{HTML}{70BF41}
\definecolor{knblue}{HTML}{51A7F9}
\definecolor{knpurple}{HTML}{B36AE2}

\def\nxlo#1{{N$^{#1}$LO}}

\def\d{{\delta}}

\def\e{{\epsilon}}

\def\l{{\lambda}}

\def\s{{\sigma}}

\def\Lbar{{\bar{L}}}
\def\lam{\lambda}

\newcommand{\be}{\begin{equation}}
\newcommand{\ee}{\end{equation}}
\newcommand{\bee}{\begin{equation*}}
\newcommand{\eee}{\end{equation*}}
\newcommand{\bea}{\begin{eqnarray}}
\newcommand{\eea}{\end{eqnarray}}
\newcommand{\beaa}{\begin{eqnarray*}}
\newcommand{\eeaa}{\end{eqnarray*}}

%% file: affiliations.tex
\newcommand{\bern}{
	Institute for Theoretical Physics, 
	Albert Einstein Center for Fundamental Physics, 
	University of Bern, Switzerland
}

\newcommand{\cmu}{
    Department of Physics, Carnegie Mellon University,
    Pittsburgh, Pennsylvania 15213, USA
}

\newcommand{\kentState}{
	Department of Physics, Kent State University, 
	Kent, OH 44242, USA
}
\newcommand{\kentucky}{
	Department of Physics and Astronomy, University of Kentucky, 
	Lexington, KY 40506, USA
}
\newcommand{\lblnsd}{
    Nuclear Science Division,
    Lawrence Berkeley National Laboratory,
	Berkeley, CA 94720, USA
}

\newcommand{\liverpoolHope}{
	School of Mathematics, Computer Science and Engineering,
	Liverpool Hope University, Hope Park, Liverpool L16 9JD, UK
}
\newcommand{\liverpool}{
	Theoretical Physics Division, Department of Mathematical Sciences, University of Liverpool, Liverpool L69 3BX, UK
}
\newcommand{\llnl}{
	Physics Division,
	Lawrence Livermore National Laboratory,
	Livermore, CA 94550, USA
}

\newcommand{\ornl}{
	National Center For Computational Sciences
	Oak Ridge National Laboratory, 
	Oak Ridge, Tennessee, USA
}
\newcommand{\ucb}{
	Department of Physics,
	University of California,
	Berkeley, California 94720, USA
	}
\newcommand{\unc}{
	Department of Physics and Astronomy,
	University of North Carolina,
	Chapel Hill, NC 27516-3255, USA
	}
\newcommand{\wm}{
	Department of Physics, The College of William \& Mary, 
	Williamsburg, VA, 23187, USA
}

%% file: introduction.tex
\textit{Introduction}: 
Lattice QCD (LQCD) regularization methods that respect chiral symmetry~\cite{Ginsparg:1981bj,Luscher:1998pqa} are highly desirable as chiral symmetry is a near exact symmetry of QCD for light quarks, it suppresses sources of large discretization errors, and it governs the mixing pattern in Standard Model (SM) and Beyond the Standard Model (BSM) matrix element calculations such as $K-\bar{K}$ mixing.
At the same time, they require a significant increase in numerical cost to utilize.  

Domain Wall Fermions (DWF)~\cite{Kaplan:1992bt,Shamir:1993zy,Furman:1994ky} are the most popular method of incorporating (approximate) chiral symmetry through the introduction of a fifth dimension of length $L_5$ while binding the left and right handed fermions at opposite ends.
A very successful DWF program with three flavors of dynamical fermions, a degenerate pair of light up and down quarks along with a strange quark ($N_f=2+1$) has been carried out for decades, see for example Refs.~\cite{RBC-UKQCD:2008mhs,RBC:2010qam,RBC:2012cbl,RBC:2014ntl}.

At finite $L_5$, an exponentially small residual chiral symmetry breaking persists that acts as a small additive mass renormalization to the quarks, $m_{\rm res} \propto e^{-L_5}$~\cite{Furman:1994ky,Blum:2000kn}. Keeping $m_{\rm res}< m_{l}=\frac{1}{2}(m_u+m_d)$ is key to maintaining good chiral symmetry, however the numerical cost of simulations grows with $L_5$. Several strategies to allow for both small $m_{\rm res}$ and $L_5$ exist in the literature \cite{Aoki:2002vt,Vranas:2006zk,Fukaya:2006vs,HotQCD:2012vvd,Hashimoto:2014gta}, and can be combined with the M\"obius Domain Wall Fermion (MDWF) prescription~\cite{Brower:2012vk}. Some of these make it more difficult for updating algorithms to change topology, and although working at coarser lattice spacing avoids this topological freezing issue this comes at the cost of not reaching fine enough lattice spacings to include dynamical charm quarks with controlled discretization effects.

In this work, we show that a balance can be struck whereby small $m_{\rm res}$ can be obtained at very small $L_5$ while topological tunneling still occurs at fine lattice spacings down to $a\approx0.05$~fm. This is made possible by performing a judicious amount of smearing for the fermions in combination with a gauge action with a weaker rectangle term than is typically used.
We scrutinize this new action with one of the most precisely determined quantities in LQCD, $F_{K^\pm}/F_{\pi^\pm}$ which is known to 0.16\% precision~\cite{FlavourLatticeAveragingGroupFLAG:2024oxs}, and is in itself a crucial ingredient in parameterizing the SM of particle physics.

A 2-3$\sigma$ deficit in the unitarity of the first row of the Cabibbo–Kobayashi–Maskawa (CKM) matrix~\cite{Cabibbo:1963yz,Kobayashi:1973fv} has persisted under increasing precision and scrutiny~\cite{FlaviaNetWorkingGrouponKaonDecays:2010lot,Moulson:2017ive,Cirigliano:2022yyo}. 
A determination of $F_{K^\pm}/F_{\pi^\pm}= F_K/F_\pi \times (1 + \frac{1}{2}\d_{SU(2)})$ is necessary to relate the ratio of $K\rightarrow\ell\nu$ to $\pi \rightarrow\ell\nu$ decays to $|V_{us}|/|V_{ud}|$, where $F_K/F_\pi$ is the ratio in the isospin limit and $\d_{SU(2)}$ is the strong isospin breaking correction.
LQCD results that non-perturbatively include $\d_{SU(2)}$~\cite{Dowdall:2013rya,Bazavov:2017lyh} are in $1\s$ agreement with the chiral perturbation theory ($\chi$PT)~\cite{Gasser:1983yg,Gasser:1984gg} estimate~\cite{Cirigliano:2011tm}.
The LQCD average~\cite{FlavourLatticeAveragingGroupFLAG:2024oxs} values of $F_{K^\pm}/F_{\pi^\pm}$ are 1.1934(19) and 1.1916(34) from $N_f=2+1+1$~\cite{Dowdall:2013rya,Bazavov:2017lyh,Miller:2020xhy,ExtendedTwistedMass:2021qui} and $N_f=2+1$~\cite{Follana:2007uv,BMW:2010xmi,RBC:2014ntl,Durr:2016ulb,QCDSFUKQCD:2016rau,CLQCD:2023sdb} flavor calculations respectively (with a new result in Ref.~\cite{Conigli:2025kan}).

Improving the precision of $F_K/F_\pi$ remains desirable as the strong isospin breaking correction uncertainty is (marginally) sub-dominant.
Further, the $N_f=2+1+1$ results are correlated and only two groups have an uncertainty $\s_{F_K/F_\pi}\simeq 0.0020$~\cite{Dowdall:2013rya,Bazavov:2017lyh} with the others having $\s_{F_K/F_\pi}\gtrsim0.0040$~\cite{Miller:2020xhy,ExtendedTwistedMass:2021qui}.
Three of these results~\cite{Dowdall:2013rya,Bazavov:2017lyh,Miller:2020xhy} use the same fermion (HISQ~\cite{Follana:2006rc}) and gauge action as well as using many of the same configurations~\cite{MILC:2010pul,MILC:2012znn}.
It is therefore highly desirable to have new results from different LQCD actions that reach the same level of precision.

Utilizing a modest amount of computing resources, we present a LQCD calculation using 30 ensembles with $N_f=2+1+1$ flavors of MDWF and the tree-level Symanzik~\cite{Symanzik:1983gh} improved gauge action, at five values of the lattice spacing ranging from $a\in[0.05,0.12]$~fm and pion masses in the range $m_\pi\in[135,460]$~MeV, with a single ensemble generated at the physical mass point.  Our final result is:
\begin{align}\label{eq:final_result}
\frac{F_K}{F_\pi} &= 1.1984(23)^s(07)^\chi(17)^a(17)^M[34]^{\rm total}\, ,
\nonumber\\
\frac{F_{K^\pm}}{F_{\pi^\pm}} &= 1.1962(34)\, ,
\end{align}
where statistical ($s$), chiral extrapolation ($\chi$), continuum extrapolation ($a$) and Model-variance ($M$) uncertainties combine for a total uncertainty of 0.28\%.
A common estimate of the strong isospin breaking corrections is applied to shift to $F_{K^\pm}/F_{\pi^\pm}$ from the FLAG convention~\cite{FlavourLatticeAveragingGroupFLAG:2024oxs} of the isospin symmetric point.
Further details of the LQCD action, action parameters, calculational results, and extrapolation analysis follow.

%% file: ensemble_details.tex
\textit{Lattice Ensemble and Measurement Details}: 
For the gauge action, we use the tree-level Symanzik-improved form defined by the rectangle coefficient $c_1=-1/12$~\cite{Symanzik:1983gh} and bare coupling strength $\beta$.
For the MDWF action, we use 8 iterations of Stout link smearing~\cite{Morningstar:2003gk} with parameter $\rho=0.125$ (comparable to the flow time of 1 as used in the mixed-action setup in \cite{Berkowitz:2017opd}, but more tractable in dynamical-fermion ensemble generation). 
A fixed number of smearing iterations for each ensemble is used such that the smearing radius vanishes as the continuum limit is approached.
The domain wall height is set to $M_5=1.0$ on all ensembles, where it was shown in the free-field limit that the oscillatory modes decouple~\cite{Syritsyn:2007mp}.
The M\"obius parameters are set under the constraint $b-c=1.0$ such that the Dirac operator is a rescaled Shamir operator. We tune $b$ and the fifth-dimensional length $L_5$ such that $\sqrt{t_0}m_\text{res}\lesssim 0.0009$ for the light quarks, i.e. $m_\text{res} \lesssim 1.5$~MeV.

\begin{table}
  \begin{tabular}{cc|cccc|c|c|c}
    \toprule
    $\beta$ & s.h & $b$ & $c$ & $L_5$ & $M_5$ & $L_\text{traj}$ & $(am_s,am_c)$ & $10^5am_\text{res} $\\
    \hline
    4.008 & a12 & 1.75 & 0.75 & 10 & 1.0 & 0.7 & (0.0725,0.8555) & 74(5) \\
    4.068 & a10 & 1.50 & 0.50 & 8  & 1.0 & 1.0 & (0.0560,0.6608) & 43(3) \\
    4.160 & a08 & 1.35 & 0.35 & 6  & 1.0 & 1.6 & (0.0415,0.4897)$^*$ & 22(3) \\
    4.238 & a06 & 1.20 & 0.20 & 4  & 1.0 & 2.6 & (0.0305,0.3599) & 39(1) \\
    4.333 & a05 & 1.16 & 0.16 & 4  & 1.0 & 4.0 & (0.0230,0.2714) & 23(0) \\
    \botrule
  \end{tabular}
  \caption{Gauge action ($\beta$) short-hand name (s.h) used later, MDWF parameters $(b,c,L_5,M_5)$, HMC trajectory length $(L_\text{traj})$, and bare strange and charm quark masses $(am_s,am_c)$ tuned for our ensembles, as well as exemplary light-quark residual mass values. The masses with a $^*$ are our preferred value but most of our boxes at this $\beta$ have $(0.0425,0.5015)$.}\label{tab:actionpars}
\end{table}

We work with a fixed strange quark mass per $\beta$ tuned to the connected part of the unphysical ``$\eta_s$'' from \cite{HPQCD:2011qwj} in $t_0$ units \cite{Luscher:2010iy}, $\phi_{\eta_s} = 8t_0m_{\eta_s}^2 = 1.989(32)$. For the charm and strange contributions in the HMC we use the Exact One Flavor algorithm \cite{Chen:2014hyy}. From this tuned strange quark we always set $am_c=11.8am_s$, which is consistent with the current $N_f=2+1$ and $N_f=2+1+1$ results in FLAG \cite{FlavourLatticeAveragingGroupFLAG:2024oxs}, allowing for a precise and accurate determination of $m_c/m_s$ in future calculations as we begin close to the physical value.  This leaves the charm quark mass always less than the Pauli-Villars mass ($am_{\rm PV}=1$), though at the coarsest $\beta=4.008$, we observe significant discretization effects for charmonia. For our choices of $\beta$ we always have $t_0/a^2 > 1$, see Table~\ref{tab:actionpars} for a list of parameters for each $\beta$.

\begin{figure}[h!]
\centering
\includegraphics[scale=0.145]{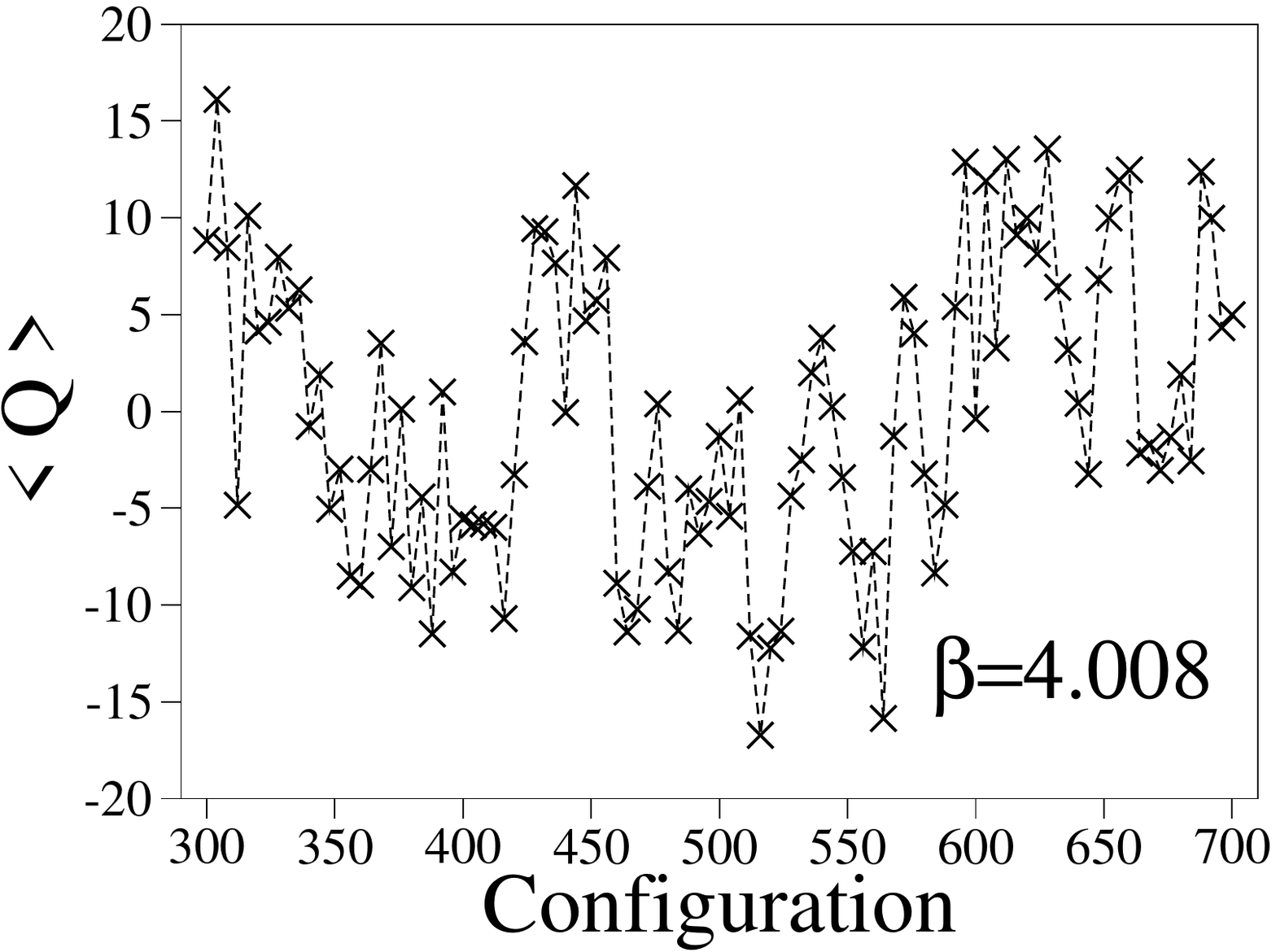}
\hspace{-12pt}
\includegraphics[scale=0.145]{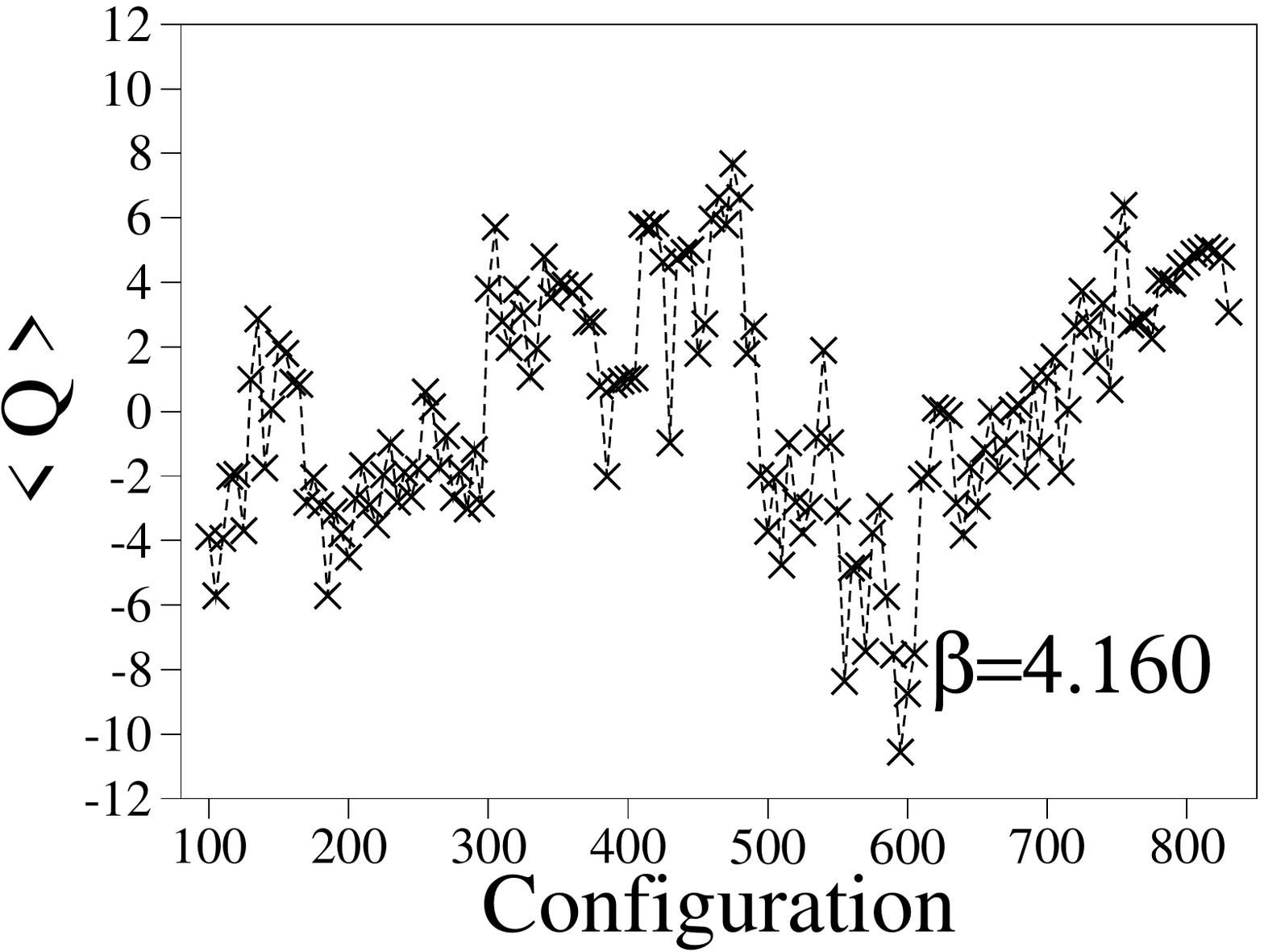}
\newline
\includegraphics[scale=0.145]{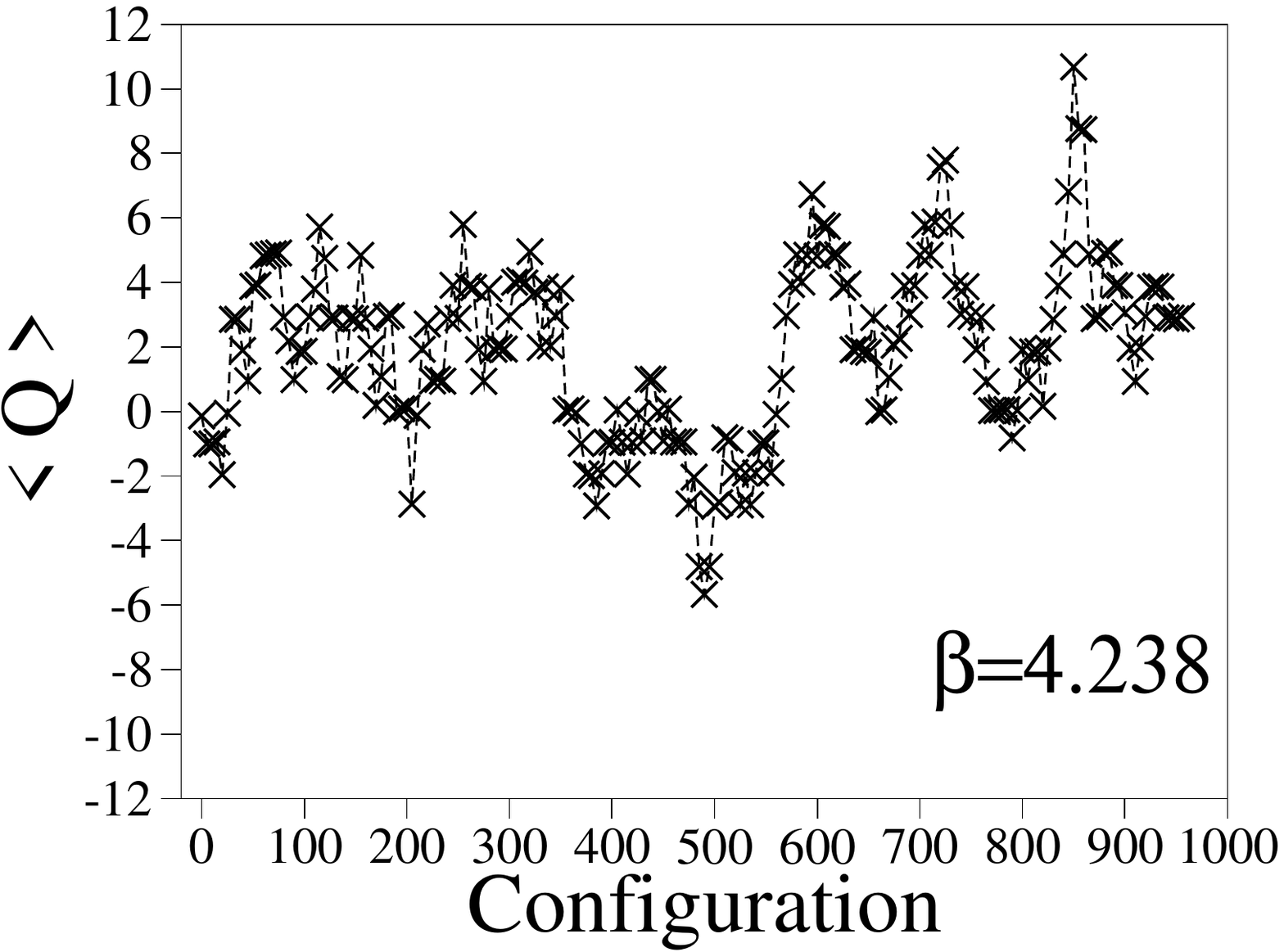}
\hspace{-12pt}
\includegraphics[scale=0.135]{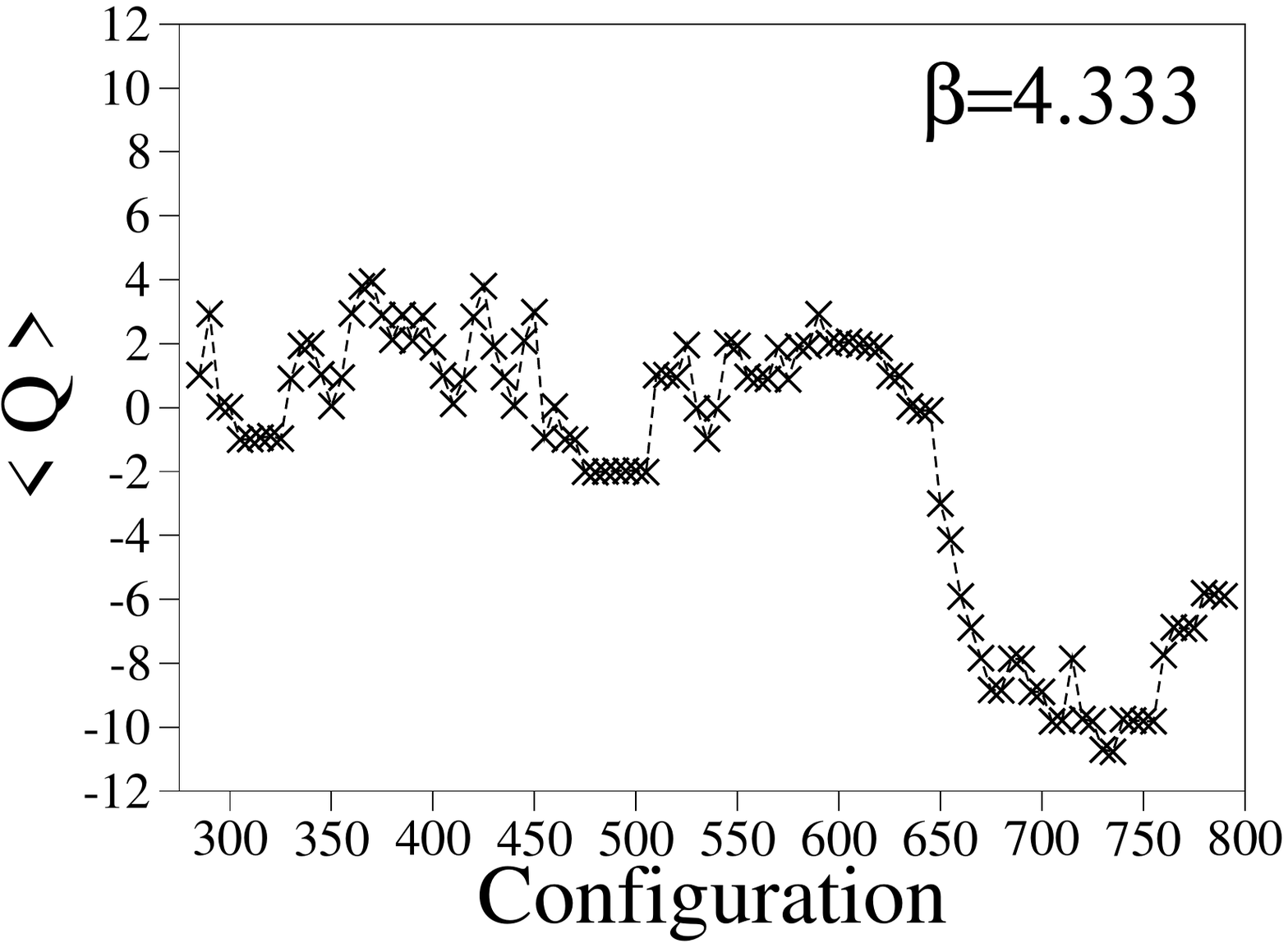}
\caption{Typical topological charge at large flow time for the coarsest and three finest $\beta$s. Lines are drawn to guide the eye.}\label{fig:topcharge}
\end{figure}

Fig.~\ref{fig:topcharge} shows our topological charge from the gauge field definition at large flow times $t\approx 1.5w_0^2$ vs. the configuration index. Only for our finest ensemble do we see an indication of slowdown with a drift to large negative topology pulling the average away from zero slightly.
We have checked that our results of $F_K/F_\pi$ are not sensitive to positive or negative topology at our current precision. 
Additional details of the ensemble generation and the topological charges are included in the End Matter
and a more extensive discussion of the new action will be provided in Ref.~\cite{cosmon:mdwf}.

In order to extract the kaon to pion decay constant ratio, we consider two-point functions of the operators
\begin{equation}
O_P = \bar{\phi}\gamma_5 \psi, \qquad O_{A_t} = \bar{\phi}\gamma_t \gamma_5 \psi,
\end{equation}
with $\psi=u$ and $\phi=s/d$. We construct diagonal two-point functions of $O_P$, as well as off-diagonal two-point functions with $O_P$ at the source and $O_{A_t}$ at the sink. As our source/sink setup is symmetric we can simultaneously fit the correlator corresponding to state ``X'' using their spectral decomposition in the asymptotic regime where the source-sink time separation $t\gg 0$:
\begin{equation}\label{eq:simPPAP}
  \begin{aligned}
    C_{P P}(t/a) &= PP( e^{-am_Xt/a} + e^{-am_X(T-t)/a}),\\ 
    C_{A_t P}(t/a) &= AP( e^{-am_Xt/a} - e^{-am_X(T-t)/a}),
  \end{aligned}
\end{equation}
with free fit parameters $A$, $P$, and $am_X$. From this fit we can directly determine the decay constant,
\begin{equation}
  aF_X = Z_A \sqrt{\frac{A^2}{am_{X}}}. \label{eq:aFx}
\end{equation}
The chiral nature of DWF means $Z_A=Z_V+O((am_\text{res})^2)$ \cite{RBC:2010qam} and so we extract $Z_V$ from pion three-point functions with a temporal vector current insertion \cite{RBC:2014ntl} and use this for $Z_A$. More details about the computation of the two- and three-point functions are given in the End Matter.

%% file: chiral_fit_form.tex
\textit{Chiral Continuum Fit Form}: The ratio of the meson decay constants can be computed in $\chi$PT~\cite{Gasser:1984gg}.
At next-to-next-to-leading order (\nxlo{2}), several non-unique choices that are equivalent can be made to express the quark/meson mass dependence.  A particularly convenient form is an analytic expression for the full two-loop result with the $\eta$ mass parameterized through the leading-order relation $m_\eta^2 = \frac{4}{3}m_K^2 -\frac{1}{3}m_\pi^2$, with the next-to-leading order (NLO) corrections to this relation accounted for in the \nxlo{2} corrections to $F_K/F_\pi$, resulting in~\cite{Ananthanarayan:2017qmx}
\begin{align}\label{eq:FKFpi_nnlo_Fpi}
    \frac{F_K}{F_\pi} &= 1
    +\frac{5}{8}\ell_\pi - \frac{1}{4}\ell_k -\frac{3}{8}\ell_\eta
    +4 \Lbar_5 (\xi_K - \xi_\pi)
\nonumber\\&\phantom{=}
    +\xi_K^2 F_F\left(\frac{m_\pi^2}{m_K^2}\right)
    +\hat{K}_1^r \lam_\pi^2
    +\hat{K}_2^r \lam_\pi \lam_K
\nonumber\\&\phantom{=}
    +\hat{K}_3^r \lam_\pi \lam_\eta
    +\hat{K}_4^r \lam_K^2
    +\hat{K}_5^r \lam_K \lam_\eta
    +\hat{K}_6^r \lam_\eta^2
\nonumber\\&\phantom{=}
    +\hat{C}_1^r \lam_\pi
    +\hat{C}_2^r \lam_K
    +\hat{C}_3^r \lam_\eta
    +\hat{C}_4^r\,.
\end{align}
Here, $\l_P = \ln (m_P^2/\mu^2)$ for $P\in\{\pi, K, \eta\}$ and
\begin{equation}
\ell_P = \xi_P \l_P = \frac{m_P^2}{(4\pi F_\pi)^2} \ln \frac{m_P^2}{\mu^2},\
\bar{L}_5 = (4\pi)^2 L_5^r(\mu)\, .
\end{equation}
In addition to the definition of $m_\eta$, which is often not computed as it requires disconnected diagrams, the use of $F_\pi$ rather than $F_0$ appearing in the definition of $\xi_P$ simplifies the analysis. This is because $F_\pi$ can be computed on each ensemble, thus avoiding the need for scale setting, which would otherwise correlate the results across all ensembles.
Scale setting is also avoided by setting the renormalization scale appearing in the logarithms as $\mu=4\pi F_\pi$~\cite{Beane:2006kx} and relating this to a scale $\mu_0=4\pi F_0$ through corrections that appear starting at \nxlo{2}~\cite{Miller:2020xhy}.

The low-energy-constants (LECs) that contribute are $L_5^r$ at next-to-leading order (NLO) as well as combinations of the other NLO LECs ($L_i^r$) and LECs appearing at \nxlo{2}, encapsulated in the $\hat{C}_i^r$ terms.  The $\hat{K}_i^r$ terms are products of $\xi_P \xi_{P^\prime}$ and numerical coefficients.  The $F_F(x)$ function captures non-analytic contributions from the double sunset graphs.  All of these coefficients are provided in Ref.~\cite{Ananthanarayan:2017qmx}.

Discretization errors are flavor blind and so they must be proportional to $(\xi_K-\xi_\pi)$ to some power, with the leading discretization corrections appearing at \nxlo{2}.
We also consider \nxlo{3} discretization corrections
\begin{align}
\delta_{a}^{\text{\nxlo{2}}} &=  A^4_s\e^2_a (\xi_K - \xi_\pi) \, ,
\quad \e_a^2 = a^2 / (4w_0^2)\, , \\
\delta_{a}^{\text{\nxlo{3}}} &= \e^2_a (\xi_K - \xi_P) ( A^6_{sK}\xi_K + A^6_{s\pi}\xi_\pi)
\nonumber\\&\phantom{=}+A^6_s\e^4_a (\xi_K - \xi_\pi) \, ,
\end{align}
where $A_i^{4,6}$ are lattice action dependent LECs characterizing discretization effects where a power counting that treats $\mathrm{O}(\e_a^2) \sim \mathrm{O}(\xi_P)$ is assumed.

The periodic finite-volume (FV) induces corrections to $F_K$ and $F_\pi$ that are characterized through modifications of the loop diagrams appearing in $\chi$PT~\cite{Gasser:1986vb,Colangelo:2004xr,Colangelo:2005gd}. At NLO in $\xi_P$, the tadpole integrals can be written as 
\begin{equation}
\ell^{\rm FV}_P = \xi_P \left[\lambda_P + 4 \sum_{|\textbf{n}|} c_n \frac{K_1(m_P L |\textbf{n}|)}{m_P L |\textbf{n}|} \right] \, ,
\end{equation}
where $c_n$ are multiplicity factors counting the number of combinations that $(n_i, n_j, n_k)$ can form a vector of length $|\textbf{n}|$ (eg. Table 1 of Ref.~\cite{Colangelo:2005gd}). $K_1(x)$ is a modified Bessel function of the second kind. 
FV corrections at \nxlo{2} have been worked out~\cite{Bijnens:2014dea} but our data is not precise enough to be sensitive to them as discussed below.

%% file: results.tex
\textit{Results}: 
To perform the extrapolation to the physical point, we closely follow the strategy described in detail in Ref.~\cite{Miller:2020xhy} and consider a variety of $\chi$PT extrapolation functions that are equivalent at \nxlo{2}, with differences starting at \nxlo{3}.
We also consider models where only the counterterms are added at \nxlo{2} (no $\ln m_P^2$ or $\ln^2 m_P^2$ terms) as well as the inclusion of the full set of \nxlo{3} counterterms or just the $\e_a^4(\xi_K-\xi_\pi)$ one. The results of $F_K/F_\pi$, and chiral parameters $\xi_\pi,\xi_K,m_\pi L,$ and $\epsilon_a^2$ necessary for this extrapolation are listed in the End Matter as well as details of these various extrapolation models.

\begin{figure}
\includegraphics[width=\columnwidth]{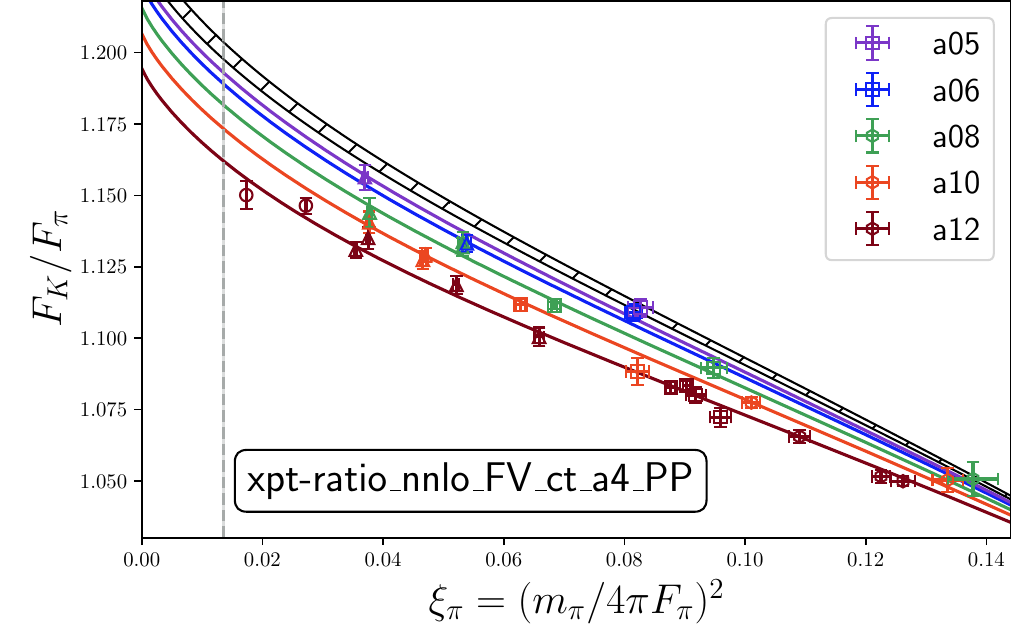}
\includegraphics[width=\columnwidth]{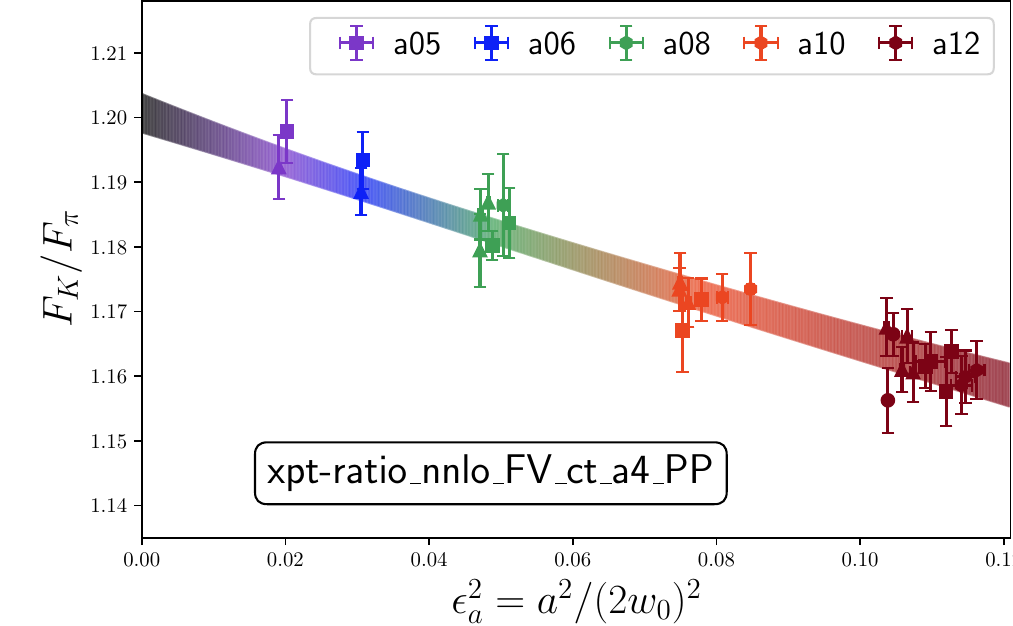}
\includegraphics[width=\columnwidth]{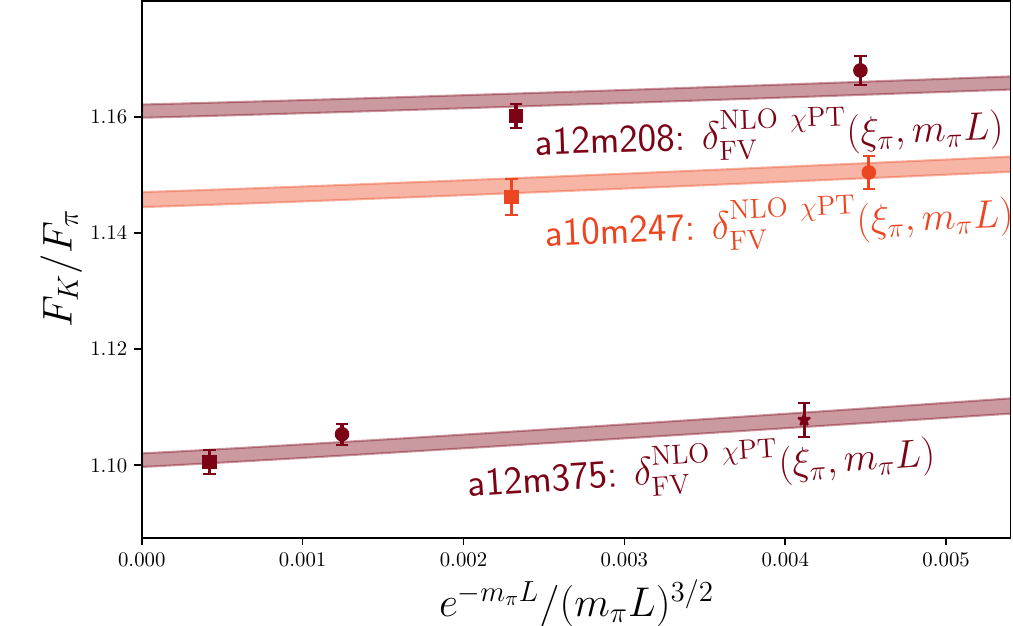}
\caption{\label{fig:exemplary_extrapolation}
Extrapolation plots from one of our highest-weight chiral-continuum fits.  The light quark mass extrapolation (top), continuum extrapolation (middle) and infinite volume extrapolation (bottom) are shown after a global analysis.
The top dashed band is the continuum, infinite volume limit.
In the middle plot, all x-variables have been shifted to the physical point except the lattice spacing.
In the bottom plot, a12m208 indicates $a\approx0.12$~fm and $m_\pi\approx208$~MeV etc.
}
\end{figure}

In total, we use 16 different models to perform this global extrapolation, with details of the model and results collected in the End Matter.
A Bayesian framework is used to perform the analysis following Ref.~\cite{Miller:2020xhy}.
All priors and posteriors are Gaussian allowing for an analytic determination of the Bayes Factor which is proportional to the relative probability of the model given a fixed set of data.  We utilize this Bayes Factor to construct weights for performing a model averaging to arrive at our final result in the isospin limit given in \eqnref{eq:final_result}.
An interesting observation from this analysis is that the non-analytic $\ln(m_\pi)$ and $\ln^2(m_\pi)$ corrections arising at \nxlo{2} are strongly disfavored to describe the results, which was also observed in Ref.~\cite{Miller:2020xhy}.

In \figref{fig:exemplary_extrapolation}, we display the extrapolation to the physical point along the three main extrapolation axes, from one of the highest weighted extrapolation models.
The top panel displays the extrapolation in $\xi_\pi$ and illustrates the span of ensembles in the light-quark mass.
The middle panel shows the continuum extrapolation.
In this figure, the global fit has been used to shift all the data to the physical values of $\xi_\pi$, $\xi_K$ and the infinite volume.
For illustrative purposes, we chose a fit that included the $\e_a^4(\xi_K-\xi_\pi)$ \nxlo{3} counterterm.  
Finally, the bottom panel displays the extrapolation to the infinite volume.
In this plot, the data are all shifted to the values of $\xi_\pi$ and $\xi_K$ corresponding to the largest volume.  The bands are the FV predictions from NLO.  Our results are not precise enough to be sensitive to higher-order FV corrections.

\begin{figure}
\includegraphics[width=\columnwidth]{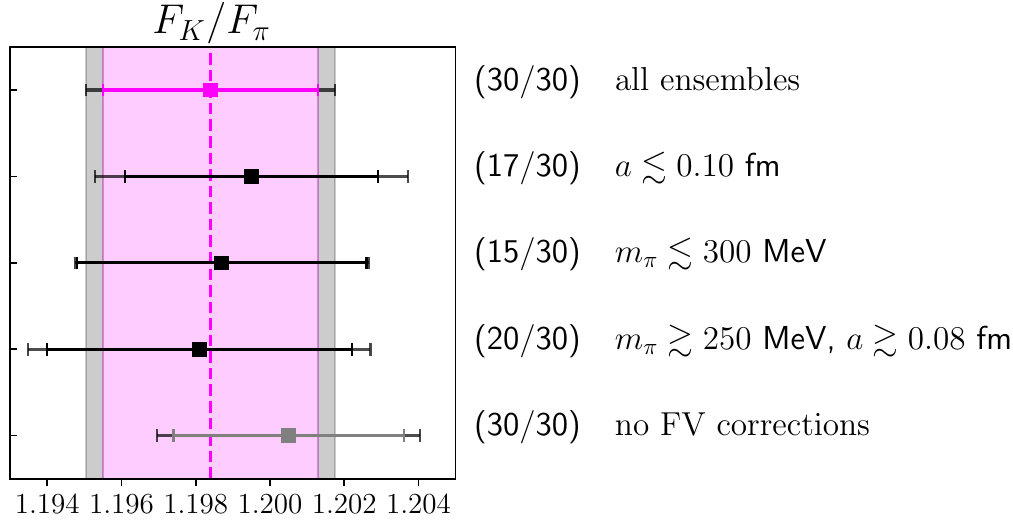}
\caption{\label{fig:data_cut}
Resulting $F_K/F_\pi$ in the isospin symmetric limit under various data cuts.
A comparison with the full data but without any FV corrections is also shown.
}\label{fig:cutdata}
\end{figure}

Most of our results come from the coarsest lattice spacing and pion masses heavier than physical, furthest from the physical point.  Therefore, it is important to check the sensitivity of our final result to various cuts in the data, which is displayed in \figref{fig:data_cut}.  Cutting all the coarsest data ($a\lesssim0.10$~fm) leads to an increase in the mean by $\sim1/3\ \s$.  Cutting all results with $m_\pi\gtrsim300$~MeV ($m_\pi\lesssim300$~MeV) barely shifts the central value.
Results from both of these data truncations are well contained within the quoted final uncertainty demonstrating that at the current precision, our results are not biased by the ensembles furthest from the physical point.

It is also instructive to cut data closest to the physical point to see how much influence these ensembles have on our final result.  We performed an analysis where the finest two lattice spacings and all ensembles with $m_\pi\lesssim250$~MeV are removed ($m_\pi\gtrsim250$~MeV, $a\gtrsim0.08$~fm).  From the growth in the final uncertainty, it is clear how valuable these finest lattice spacing and near physical pion mass results are for improving the precision.

We repeat the main analysis without FV corrections.  The result is still compatible with our final result, further demonstrating our results are not sensitive to higher-order FV corrections with the present level of precision.  While compatible, turning off the NLO FV corrections leads to $\chi^2_{aug}\geq2.13$ and overall very poor fit quality.

For the \nxlo{2} extrapolation models, we list in \tabref{tab:L5} the resulting value of $L_5$ determined.  The fit values are evolved from $\mu_0=4\pi F_0$ to the standard scale $\mu=770$~MeV using the NLO scale dependence~\cite{Gasser:1984gg}.  Details of these models are given in the End Matter.

\begin{table}
\caption{\label{tab:L5}
Resulting value of $10^3 L_5(\mu=770\ \textrm{MeV})$ from select fit models.
The NLO $\mu$ dependence was used to evolve them to the standard scale at the rho mass.}
\begin{ruledtabular}
\begin{tabular}{rccl}
model& \multicolumn{2}{c}{$10^3 L_5(\mu=770\ \textrm{MeV})$}& model \\
\hline
nnlo        & -0.11(38)& 0.07(38)& nnlo\_ratio\\
nnlo\_a4    & -0.10(38)& 0.08(38)& nnlo\_ratio\_a4\\
nnlo\_ct    & -0.04(11)& 0.34(16)& nnlo\_ratio\_ct\\
nnlo\_ct\_a4& -0.01(12)& 0.37(16)& nnlo\_ratio\_ct\_a4\\
\end{tabular}
\end{ruledtabular}
\end{table}

In order to correct for strong isospin breaking corrections, we use the FLAG definition of the isospin symmetric point~\cite{FlavourLatticeAveragingGroupFLAG:2024oxs} (the ``Edinburgh Consensus'').
Using a definition of the isospin limit suggested in Ref.~\cite{Cirigliano:2011tm} leads to a difference from \eqnref{eq:final_result} of 0.0003, or 10\% of the statistical uncertainty.
We apply a strong isospin breaking correction following Ref.~\cite{Miller:2020xhy} which builds upon that of Ref.~\cite{Cirigliano:2011tm}, arriving at
\begin{equation}
0.5 \times \d_{SU(2)} = -0.00216(54)\, ,
\end{equation}
and our prediction for $F_{K^\pm} / F_{\pi^\pm}$ in \eqnref{eq:final_result}.

%% file: conclusions.tex
\textit{Conclusions}: We have demonstrated that dynamical, $N_f=2+1+1$ MDWF ensembles can be used for high-precision physics, providing results consistent with the world average for the benchmark quantity $F_K/F_\pi$ at the 0.3\% level of precision. This illustrates that a dynamical, lattice simulation with four flavors of chiral fermions can be performed with relatively modest computational resources. 
The results presented here, ensemble generation and correlation functions, were generated with a remarkably small $O(40\rm{k})$ Perlmutter-GPU node hours.  
We achieved this by using the Symanzik gauge action in combination with the M\"obius Domain Wall Fermion prescription and judicious smearing, roughly comparable to working at a flow time of 1. Without the smearing it would not have been possible to work at the small values of $L_5$ we do, and would have likely made simulations at the coarse lattice spacing intractable.

Our results indicate no significant problems with simulating this action down to $O(0.05)$ fm with periodic boundary conditions for the gauge fields, and we observe small $m_\text{res}$ even with very short fifth dimension lengths of only 4 lattice units. There is no fundamental problem with generating ensembles at the physical pion mass point using this action and further ensembles at the physical point are planned.

\begin{figure}[h!]
  \includegraphics[scale=0.28]{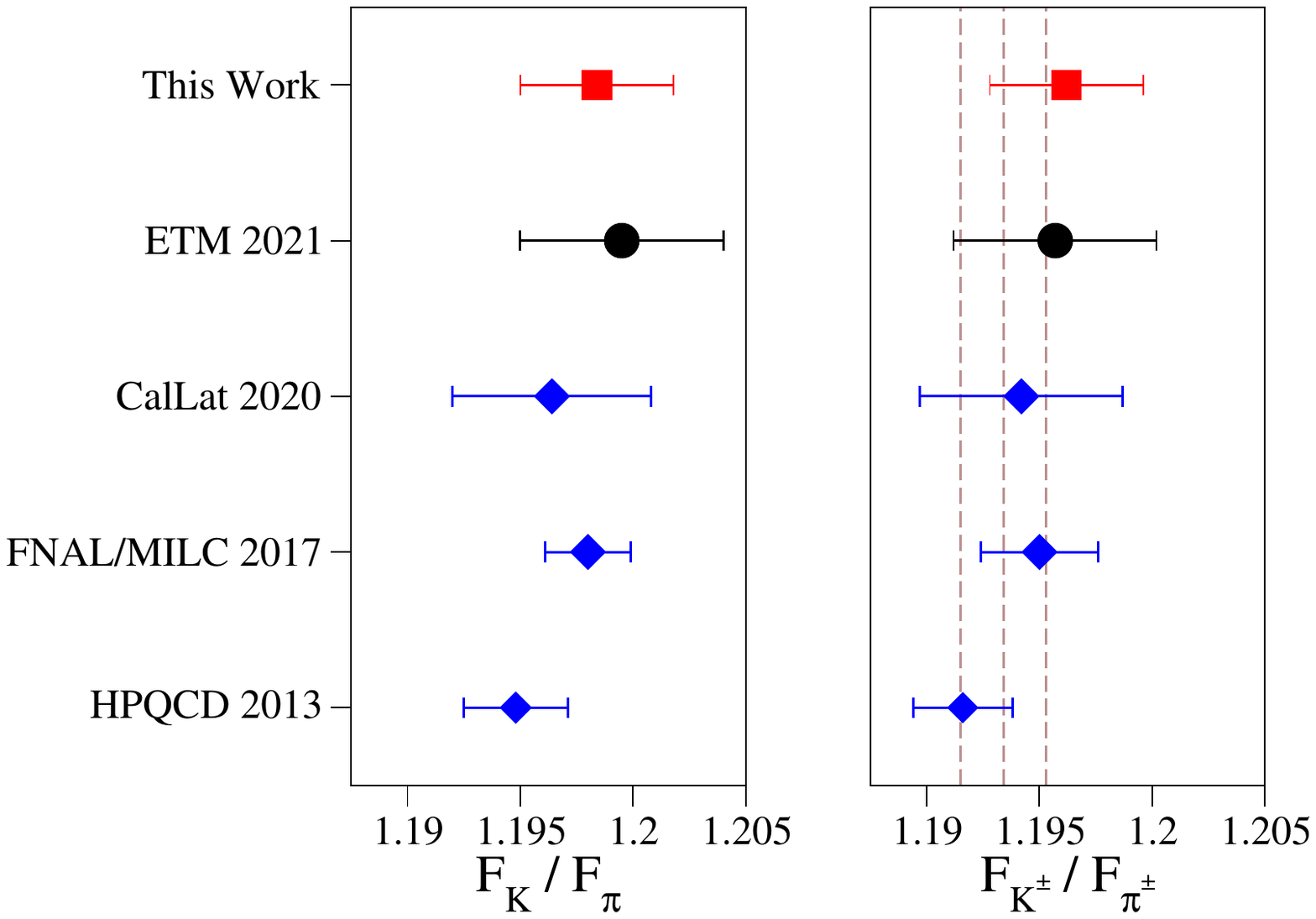}
  \caption{A comparison of our determinations of $F_K/F_\pi$ and $F_{K^\pm} / F_{\pi^\pm}$ with the most-precise $N_f=2+1+1$ results that enter the FLAG 2024 average (shown as a brown dotted line) \cite{Dowdall:2013rya,Bazavov:2017lyh,Berkowitz:2017opd,ExtendedTwistedMass:2021qui}, statistical and systematic errors have been added in quadrature. Red squares indicate our dynamical MDWF in the sea, black circles are twisted-mass Wilson, blue diamonds have rooted-staggered quarks.}\label{fig:nf2p1p1_lit}
\end{figure}

Our final result, \eqnref{eq:final_result}, is obtained from 30 ensembles with 5 different lattice spacings and a large range of pion masses, down to physical.
Fig.~\ref{fig:nf2p1p1_lit} illustrates our compatibility for $F_K/F_\pi$ and $F_{K^\pm}/F_{\pi^\pm}$ with the other LQCD results. Blue points~\cite{Dowdall:2013rya,Bazavov:2017lyh,Berkowitz:2017opd} indicate that rooted-staggered quarks were used in the sea, sharing most of the configurations between the three determinations. The black circle is the determination from ETMC~\cite{ExtendedTwistedMass:2021qui} and the red square is the present calculation, while the dashed line indicates the 2024 FLAG average~\cite{FlavourLatticeAveragingGroupFLAG:2024oxs}. The HPQCD determination introduces some tension in the average and more recent determinations, including this one, indicate a slightly larger value of $F_{K^\pm}/F_{\pi^\pm}$.
By providing a third completely independent determination with a different lattice action, we add further credence to the remarkably consistent high-precision 4-flavor global lattice average.

Adding our current result, \eqnref{eq:final_result}, to the $N_f=2+1+1$ FLAG average~\cite{FlavourLatticeAveragingGroupFLAG:2024oxs} reduces the quoted uncertainty by a modest 10\%, having minimal impact on the first row CKM unitarity test.
With a moderate increase in computing resources, we will be able to generate two additional ensembles at the physical pion mass with finer lattice spacings, significantly reducing the uncertainty of our final result to match that of Refs.~\cite{Dowdall:2013rya,Bazavov:2017lyh}.
With such a result, the theoretical uncertainty in determining $|V_{us}|/|V_{ud}|$ would become sub-dominant to the experimental uncertainties for all quantities~\cite{Eckhause:1965zz, Nordberg:1967zz, Auerbach:1967nip, Ayers:1971kz, Chiang:1972rp, Dunaitsev:1972st, Vaisenberg:1976tz, Bryman:1985bv, Britton:1992pg, Usher:1992pz, Czapek:1993kc, Numao:1995qf, Koptev:1995je, KLOE:2005xes, KLOE:2008tnn, PiENu:2015seu} used in global averages of first row unitarity tests~\cite{FlaviaNetWorkingGrouponKaonDecays:2010lot,Moulson:2017ive,Cirigliano:2022yyo}.
Moreover, it would push the community fully into the precision where non-perturbative isospin breaking corrections from $m_d-m_u$ and QED radiative corrections are required for which there are already a few LQCD results~\cite{Giusti:2017dwk,DiCarlo:2019thl,Boyle:2022lsi}.

The new ensembles we have generated and these results similarly pave the way for other crucial quantities to be computed with sub-percent precision, such as the $K_{\ell 3}$ ($K\rightarrow\pi\ell\nu$) form factor which also plays a significant role in first-row CKM unitarity tests, as well as a broad set of low-energy nuclear and particle physics quantities that go into precision tests of the SM, fundamental symmetry tests and general hadron interactions and reactions.

%% file: end_matter.tex

\textbf{END MATTER}\vspace{-2em}
\appendix
\section{Details of the lattice computation}

\begin{figure}[h]
\includegraphics[width=\columnwidth]{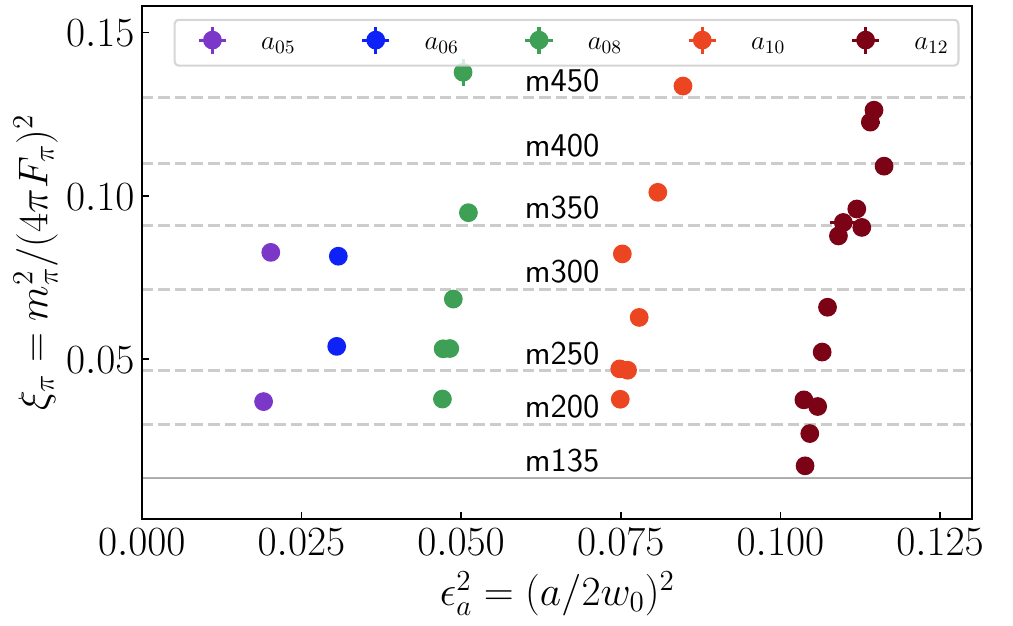}
\caption{\label{fig:landscape}
The landscape of pion mass and lattice spacing of our ensembles, using $\xi_\pi$ and $\e_a^2$ as proxies for these variables.}
\end{figure}

When generating the ensembles, a reference Molecular-Dynamics integration length of 1 was used on the $\beta=4.068$ ensembles. We then interpret the fictitious Molecular-Dynamics integration time as a quantity to be kept constant with a physical length scale for which $t_0$ was used.  This resulted in integration lengths of $0.75, 1.6, 2.6$, and $4.0$ on the $\beta=4.008, 4.160, 4.238$ and $4.333$ ensembles respectively.
To begin exploring the landscape, lattices were generated with $m_\pi L \approx 3.5$.
In certain places of interest in the space of $\beta$ and $L$, new boxes with larger $m_\pi L$ and/or larger $L/\sqrt{t_0}$ were generated. For each ensemble we have $\geq 100$ independent thermalised configurations. A comprehensive list of all relevant information for each ensemble is given in Table~\ref{tab:input_params}.
\figref{fig:landscape} shows the landscape of our ensembles in terms of pion mass and lattice spacing.

In order to extract the ratio $F_{K}/F_{\pi}$, we compute two-point functions $C_{PP}$ and $C_{A_t P}$, where $P$ refers to the pseudoscalar and $A_t$ to the temporal component of the axial vector current. We use the one-end-trick \cite{McNeile:2006bz,ETM:2008zte} and compute the quark propagators by inverting the Dirac operator on $Z_2\times Z_2$ stochastic (Z2PSWall \cite{Boyle:2008rh}) wall sources. A small number of high precision solves at some randomly-selected initial reference time position and then any other exact solves maximally-displaced in time from this initial point are used. To boost statistics, several evenly-spaced low-precision solves are performed at many different temporal translations on the same configuration, bias-correcting them with the few (often 1 or 2) high-precision determinations, as is the prescription for the Truncated Solver Method \cite{Bali:2009hu}. Because we are performing many solves on the same configuration, we find exact deflation in the single-precision part of our mixed-precision Conjugate Gradient solves to be beneficial for the smaller boxes ($L\leq 40$). The two-point functions $C_{PP}$ and $C_{A_t P}$ are then simultaneously fit the single-state ans\"atze defined in Eq.~\eqref{eq:simPPAP} to extract the ratio $F_{K}/F_{\pi}$ as well as the masses and overlap factors using Eq.~\eqref{eq:aFx}, as described in the main text. The final results are included in Table~\ref{tab:input_params}.

In addition to the above, we also need to determine the axial-vector current renormalization constant $Z_A=Z_V$ for chiral fermions, which is used to determine the quantities $\xi_\pi$ and $\xi_K$. To that end, we construct the three-point function $C_{P V_t P}(t,p=0)$, where $V_t=\bar{d}\gamma_4 u$ is the temporal vector current, using the sequential source-through-the-sink method and performing the thermal correction of \cite{Green:2017keo}. $Z_V$ is equal to the inverse of the ground state matrix element contributing to this three-point function, which can be easily extracted by a fit to a plateau in the large Euclidean operator insertion time region where the source and sink are maximally separated by $T/2$.

\section{Details of the extrapolation}

There are many equivalent forms for $F_K/F_\pi$ at \nxlo{2}.
First, the denominator is often Taylor expanded,
\begin{align}\label{eq:fkfpi_fully_expanded}
\frac{F_K}{F_\pi} &= \frac{1 + \d F_K^{\rm NLO} + \d F_K^\text{\nxlo{2}} +\cdots}
    {1 + \d F_\pi^{\rm NLO} + \d F_\pi^\text{\nxlo{2}}+\cdots} ,
\nonumber\\&= 1 + \d F_K^{\rm NLO} - \d F_\pi^{\rm NLO}
    + \d F_K^\text{\nxlo{2}} -\d F_\pi^\text{\nxlo{2}}
\nonumber\\&\phantom{=1}
    - \d F_K^{\rm NLO} \d F_\pi^{\rm NLO}
    + (\d F_\pi^{\rm NLO})^2
    +\cdots
\end{align}
producing an expression that is common in the literature~\cite{Amoros:1999dp,Bijnens:2005ae,Bijnens:2006jv,Bijnens:2014dea,Bijnens:2015dra,Ananthanarayan:2017qmx}.
At the same time, the NLO terms are large enough, $\mathrm{O}(20\%)$, that the difference in the ratio and Taylor expanded form is larger than the statistical uncertainty of the results.  Thus, one could consider an expression that is equivalent at \nxlo{2}
\begin{equation}
\label{eq:fkfpi_nlo_ratio}
\frac{F_K}{F_\pi} = \frac{1 + \d F_K^{\rm NLO}}{1 + \d F_\pi^{\rm NLO}}
    + \d F_K^\text{\nxlo{2}} - \d F_\pi^\text{\nxlo{2}}
    +\cdots
\end{equation}
where in both expressions, the $\cdots$ represent terms at \nxlo{3} and higher.
In our extrapolation analysis, the key ``\_ratio'' indicates the use of \eqnref{eq:fkfpi_nlo_ratio} while its exclusion indicates \eqnref{eq:fkfpi_fully_expanded}.
The \nxlo{2} fits with the label ``\_ct'' means that only the first line of \eqnref{eq:FKFpi_nnlo_Fpi} and the $\hat{C}_4^r=(\xi_K-\xi_\pi)(A_K^4 \xi_K + A_\pi^4 \xi_\pi)$ terms are included.
We explore adding the full \nxlo{3} counterterms (nnnlo) as well as only including $\e_a^4$ discretization errors from \nxlo{3} (\_a4).
The full list of 16 models considered is provided in \tabref{tab:extrapolation_fits} along with the final result for $F_K/F_\pi$.

Our results show relatively large discretization effects in comparison to \cite{Dowdall:2013rya,Bazavov:2017lyh} and \cite{ExtendedTwistedMass:2021qui} and more in line with the mixed-action determination of \cite{Miller:2020xhy}, although with less-pronounced $a^4$ effects. For the HISQ valence determination this difference is unsurprising as $O(a^2)$ corrections are perturbatively removed. Fortunately, having 5 lattice spacings allows us to control our continuum extrapolation. Whether we use $a^2/w_0^2$ or $a^2/t_0$ does not matter, and we arrive at the same continuum result well within errors upon removing our coarsest lattice, indicating that our continuum extrapolation is reliable.

\begingroup \squeezetable
\begin{table*}
\caption{\label{tab:input_params}
Volume and light quark mass in lattice units for the ensembles used in this work.
Additionally, the gradient flow scales ($t_0$, $w_0$), topological charge ($Q$) and their respective autocorrelation lengths are tabulated.
The derived values of $F_K/F_\pi$, $\xi_\pi = m_\pi^2/(4\pi F_\pi)^2$, $\xi_K=m_K^2/(4\pi F_\pi)^2$, $m_\pi L$ and $\epsilon_a^2 = a^2 / (4w_0^2)$ are listed as well as the number of independent configurations used $N_\text{cfg}$ and the number of time-translated sloppy solves $N_\text{src}$.
}
\begin{ruledtabular}
\begin{tabular}{cccccc|cccccc}
$\frac{L}{a}^3\times \frac{T}{a}$ & $am_l$ & $\sqrt{t_0}/a$ & $w_0/a$ & $Q$ & $\tau_{\sqrt{t_0}},\tau_{w_0},\tau_Q$ &
    $F_K/F_\pi$ & $\xi_\pi$ & $\xi_K$ & $m_\pi L$& $\epsilon_a^2$& $N_\text{cfg}\times N_\text{src}$\\
\hline
\multicolumn{12}{c}{$\beta=4.008$, $a\approx0.12$~fm}\\
\hline
$16^3\times 32$ & 0.0300 & 1.1961(31) & 1.4766(82)  &  0.3(2) & (16,16,16)&
    1.0800(20) & 0.1262(19) & 0.2106(28) & 4.424(18)& 0.1147(13) & $173 \times 32$ \\
$16^3\times 32$ & 0.0270 & 1.1946(20) & 1.4766(82)  &  0.4(3) & (16,24,24)&
    1.0898(24)& 0.1225(15)& 0.2178(24)& 4.255(16)& 0.1141(15) & $179 \times 32$ \\
$16^3\times 32$ & 0.0235 & 1.1891(20) & 1.4766(82)  & -0.2(3) & (16,16,16)&
    1.1012(25) & 0.1091(17)& 0.2150(27)& 3.979(18) & 0.1162(11) & $118 \times 32$\\
$16^3\times 32$ & 0.0195 & 1.2016(40) & 1.5086(132) &  0.2(3) & (24,24,16)&
    1.1109(33)& 0.0918(17)& 0.2070(30)& 3.580(19)& 0.1098(20) & $106 \times 32$ \\
$20^3\times 48$ & 0.0195 & 1.1969(13) & 1.5086(132) &  0.2(3) & (24,24,16)&
    1.1061(17)& 0.09031(66)& 0.2048(12)& 4.449(11)& 0.11275(67) & $204 \times 48$ \\
$24^3\times 48$ & 0.0195 & 1.2040(40) & 1.5138(57) &  0.4(5) & (16,16,8)&
    1.1007(20)& 0.08770(76)& 0.2003(16)& 5.276(11)& 0.10909(85) & $95 \times 32$ \\
$16^3\times 32$ & 0.0190 & 1.1964(13) & 1.4943(39) & -0.1(2) & (8,8,12)&
    1.1198(32)& 0.0959(17) & 0.2221(30)& 3.550(20)& 0.11197(59) & $235 \times 32$\\
$20^3\times 40$ & 0.0120 & 1.2064(7)  & 1.5258(27) &  0.6(1) & (8,16,8)&
    1.1382(30)& 0.06587(76)& 0.2149(19)& 3.541(12)& 0.10738(39) & $353 \times 20$\\
$24^3\times 48$ & 0.0090 & 1.2075(17) & 1.5319(46) & -0.3(6) & (16,16,16)&
    1.1520(26)& 0.05215(49)& 0.2137(16)& 3.7086(93)& 0.10654(64) & $116 \times 48$ \\
$28^3\times 64$ & 0.0057 & 1.2138(6)  & 1.5528(28) & -0.6(3) & (12,24,8)&
    1.1743(28)& 0.03752(30)& 0.2204(14)& 3.5223(73)& 0.10368(38) & $176 \times 64$\\
$32^3\times 64$ & 0.0057 & 1.2073(6)  & 1.5639(20) &  0.6(5) & (16,16,8)&
    1.1606(21)& 0.03548(23)& 0.2118(10)& 3.9883(75)& 0.10584(28) & $154 \times 64$\\
$32^3\times 64$ & 0.0040 & 1.2090(8)  & 1.5460(22) &  0.3(4) & (16,16,16)&
    1.1792(20)& 0.02720(20)& 0.2147(11)& 3.4247(72)& 0.10460(31) & $157 \times 64$ \\
$48^3\times 96$ & 0.0020 & 1.2120(3)  & 1.5514(11) & -0.1(10) & (12,24,12)&
    1.1894(36)& 0.01733(11)& 0.22251(99)& 3.9631(74)& 0.10386(21) & $101 \times 48$ \\
\hline
\multicolumn{12}{c}{$\beta=4.068$, $a\approx0.10$~fm}\\
\hline
\underline{$16^3\times 32$} & 0.0220 & 1.4076(19) & 1.7175(48)  & -0.1(1) & (12,12,24)&
    1.0995(40)& 0.1336(25)& 0.2236(36)& 3.745(22)& 0.08475(47) & $483 \times 32$ \\
$20^3\times 40$ & 0.0185 & 1.4223(21) & 1.7589(75)  & -0.1(3) & (16,16,40)&
    1.0963(25)& 0.1010(15)& 0.1975(24)& 4.074(18)& 0.08081(75) & $128 \times 40$ \\
$20^3\times 40$ & 0.0135 & 1.4392(30) & 1.8227(106) & -0.5(3) & (16,16,16)&
    1.1259(51)& 0.0822(19)& 0.2090(37)& 3.430(21)& 0.07525(90) & $150 \times 40$ \\ 
$24^3\times 48$ & 0.0100 & 1.4319(17) & 1.7915(77)  & -0.7(5) & (20,40,40)&
    1.1286(26)& 0.06275(79)& 0.1938(18)& 3.628(14)& 0.07789(70) & $174 \times 24$ \\
$28^3\times 64$ & 0.0067 & 1.4439(11) & 1.8276(27)  & -0.5(7) & (12,12,24)&
    1.1495(27)& 0.04700(47)& 0.1979(13)& 3.514(12)& 0.07484(22) & $115 \times 64$ \\
$32^3\times 64$ & 0.0067 & 1.4370(12) & 1.8128(38)  & -0.5(3) & (16,16,8)&
    1.1464(31)& 0.04659(46)& 0.1988(15)& 3.998(13)& 0.07607(32) & $115 \times 32$ \\    
$32^3\times 64$ & 0.0050 & 1.4430(11) & 1.8266(30)  &  0.5(4) & (12,12,12)&
    1.1646(39)& 0.03770(48)& 0.2026(19)& 3.497(12)& 0.07493(25) & $100 \times 32$ \\
\hline
\multicolumn{12}{c}{$\beta=4.160$, $a\approx0.08$~fm}\\
\hline
\underline{$20^3\times 40$}   & 0.0170 & 1.8136(51) & 2.2291(146) & 0.4(2) & (20,20,30)&
    1.1242(58)& 0.1378(41)& 0.2396(62)& 3.626(28)& 0.05031(67) & $233 \times 20$ \\
$24^3\times 48$   & 0.0165 & 1.8097(46) & 2.2108(118) & 0.0(5) & (20,20,30)&
    1.1195(44)& 0.0948(22)& 0.2049(38)& 3.705(21)& 0.05115(58) & $112 \times 24$ \\  
$32^3\times 64$   & 0.0090 & 1.8268(14) & 2.2637(35)  & -0.6(3) & (15,15,30)&
    1.1215(14)& 0.06837(45)& 0.18874(98)& 4.1297(83)& 0.04879(15) & $307 \times 64$ \\
$32^3\times 64$   & 0.0060 & 1.8318(24) & 2.2767(55)  &  0.1(5) & (15,15,25)&
    1.1580(37)& 0.05323(80)& 0.2003(22)& 3.453(13)& 0.04823(23) & $106 \times 32$ \\
$32^3\times 64^*$ & 0.0060 & 1.8398(29) & 2.3017(73)  &  0.6(7) & (20,20,40)&
    1.1538(32)& 0.05315(65)& 0.1972(18)& 3.417(12)& 0.04719(31) & $101 \times 64$ \\
$40^3\times 80$   & 0.0039 & 1.8383(12) & 2.2304(28)  &  0.1(8) & (15,15,40)&
    1.1703(51)& 0.03777(38)& 0.2028(15)& 3.4971(96)& 0.04708(12) & $147 \times 20$ \\
\hline
\multicolumn{12}{c}{$\beta=4.238$, $a\approx0.06$~fm}\\
\hline
$32^3\times 64$ & 0.0070 & 2.3142(49) & 2.8498(164) & 0.4(3) & (30,40,60)& 
    1.1379(34)& 0.0815(14)& 0.2023(27)& 3.491(15)& 0.03078(38) & $172 \times 64$\\
$40^3\times 80$ & 0.0045 & 2.3173(20) & 2.8617(76)  & 1.5(5) & (10,20,30)&
    1.1453(32)& 0.05389(67)& 0.1879(18)& 3.508(13)& 0.03053(17) & $149 \times 40$\\
\hline
\multicolumn{12}{c}{$\beta=4.333$, $a\approx0.05$~fm}\\
\hline
$40^3\times 80$  & 0.0060 & 2.8964(55) & 3.5177(132) &  0.8(8)  & (30,30,250)&
    1.1225(43)& 0.0827(21)& 0.1886(36)& 3.675(26)& 0.02020(15) & $159 \times 20$ \\
$64^3\times 128$ & 0.0021 & 2.9334(23) & 3.6199(77)  & -1.3(12) & (15,20,40)&   
    1.1704(46)& 0.03699(51)& 0.1914(20)& 3.583(15)& 0.019078(88) & $102 \times 16 $\\
\end{tabular}
\end{ruledtabular}

\end{table*}
\endgroup

\begin{table*}
\caption{\label{tab:extrapolation_fits}
Fit results to full data set.  The augmented $\chi^2_{aug}$ is the usual $\chi^2$ plus a sum over contributions from the priors.}
\begin{ruledtabular}
\begin{tabular}{rcccccccccc}
                model & expansion& include $\ln(m_\pi)$& include \nxlo{3}& include \nxlo{3}& chi2/dof &   $Q$ &  logGBF& weight& $F_K/F_\pi$\\
                      & form& @\nxlo{2}? & $\e_a^4(\xi_K-\xi_\pi)$?& $\e_a^2(\xi_K-\xi_\pi)\xi_P$?\\
\hline
      nnlo\_ratio\_ct &\eqnref{eq:fkfpi_nlo_ratio} &\xmark &\xmark &\xmark & 1.219   &  0.190&  123.409&  0.205&  1.1989(17)\\
  nnlo\_ratio\_ct\_a4 &\eqnref{eq:fkfpi_nlo_ratio} &\xmark &\cmark &\xmark &  1.201   &  0.207&  123.393&  0.202&  1.2007(30)\\
 nnnlo\_ratio\_ct\_a4 &\eqnref{eq:fkfpi_nlo_ratio} &\xmark &\cmark &\cmark &  1.167   &  0.242&  123.049&  0.143&  1.1989(39)\\
     nnnlo\_ratio\_ct &\eqnref{eq:fkfpi_nlo_ratio} &\xmark &\xmark &\xmark &  1.194   &  0.214&  122.912&  0.125&  1.1971(33)\\
         nnlo\_ct\_a4 &\eqnref{eq:fkfpi_fully_expanded}&\xmark &\cmark&\xmark&  1.207   &  0.201&  122.701&  0.101&  1.1980(28)\\
             nnlo\_ct &\eqnref{eq:fkfpi_fully_expanded}&\xmark&\xmark&\xmark&  1.235   &  0.176&  122.572&  0.089&  1.1959(16)\\
        nnnlo\_ct\_a4 &\eqnref{eq:fkfpi_fully_expanded}&\xmark&\cmark&\xmark&  1.175   &  0.234&  122.306&  0.068&  1.1970(37)\\
            nnnlo\_ct &\eqnref{eq:fkfpi_fully_expanded}&\xmark&\xmark&\cmark&  1.209   &  0.200&  122.058&  0.053&  1.1950(32)\\
            nnnlo\_a4 &\eqnref{eq:fkfpi_fully_expanded}&\cmark&\cmark&\xmark&  1.270   &  0.147&  119.476&  0.004&  1.2006(39)\\
                nnnlo &\eqnref{eq:fkfpi_fully_expanded}&\cmark&\cmark&\cmark&  1.300   &  0.126&  119.278&  0.003&  1.1988(34)\\
     nnnlo\_ratio\_a4 &\eqnref{eq:fkfpi_nlo_ratio}&\cmark&\cmark&\xmark&  1.335   &  0.104&  118.798&  0.002&  1.2002(39)\\
         nnnlo\_ratio &\eqnref{eq:fkfpi_nlo_ratio}&\cmark&\cmark&\cmark&  1.362   &  0.089&  118.652&  0.002&  1.1985(34)\\
             nnlo\_a4 &\eqnref{eq:fkfpi_fully_expanded}&\cmark&\cmark&\xmark&  1.353   &  0.094&  118.638&  0.002&  1.2029(32)\\
                 nnlo &\eqnref{eq:fkfpi_fully_expanded}&\cmark&\xmark&\xmark&  1.375   &  0.083&  118.561&  0.002&  1.2011(23)\\
      nnlo\_ratio\_a4 &\eqnref{eq:fkfpi_nlo_ratio}&\cmark&\cmark&\xmark&  1.444   &  0.055&  117.446&  0.001&  1.2027(32)\\
          nnlo\_ratio &\eqnref{eq:fkfpi_nlo_ratio}&\cmark&\xmark&\xmark&  1.463   &  0.049&  117.398&  0.001&  1.2010(23)\\
\hline
Bayes average &&&&&&&&& 1.1984(29)(17)
\end{tabular}
\end{ruledtabular}
\end{table*}